\documentclass[reprint, 
groupedaddress,
superscriptaddress,
longbibliography,
amsmath,amssymb,
aps,
prl,
]{revtex4-2}
\usepackage[colorlinks = true,]{hyperref}
\usepackage{float}
\usepackage{graphicx}
\usepackage{dcolumn}
\usepackage{bm}
\usepackage{xcolor}
\usepackage{braket}
\usepackage{soul}

\begin{document}

\preprint{APS/123-QED}

\title{Dissipative Phase Transition in the Two-Photon Dicke Model}

\author{Aanal Jayesh Shah}
\affiliation{Department of Physics and Astronomy, Purdue University, West Lafayette, IN 47906, USA}
\email{shah601@purdue.edu}
\author{Peter Kirton}
\affiliation{Department of Physics and SUPA, University of Strathclyde, Glasgow, G4 0NG, United Kingdom}
\email{peter.kirton@strath.ac.uk}
\author{Simone Felicetti}
\affiliation{Institute for Complex Systems, National Research Council (ISC-CNR), Via dei Taurini 19, 00185 Rome, Italy}
\affiliation{Physics Department, Sapienza University, P.le A. Moro 2, 00185 Rome, Italy}
\email{simone.felicetti@cnr.it}
\author{Hadiseh Alaeian}
\affiliation{Department of Physics and Astronomy, Purdue University, West Lafayette, IN 47906, USA}
\affiliation{Elmore Family School of Electrical and Computer Engineering, Purdue University, West Lafayette, IN 47906, USA}
\email{halaeian@purdue.edu}

\date{\today}

\begin{abstract}
We explore the dissipative phase transition of the two-photon Dicke model, a topic that has garnered significant attention recently. Our analysis reveals that while single-photon loss does not stabilize the intrinsic instability in the model, the inclusion of two-photon loss restores stability, leading to the emergence of superradiant states which coexist with the normal vacuum states. Using a second-order cumulant expansion for the photons, we derive an analytical description of the system in the thermodynamic limit which agrees well with the exact calculation results. Additionally, we present the Wigner function for the system, shedding light on the breaking of the $Z_4$-symmetry inherent in the model. These findings offer valuable insights into stabilization mechanisms in open quantum systems and pave the way for exploring complex nonlinear dynamics in two-photon Dicke models.
\end{abstract}

\maketitle

\emph{Introduction -- }The Dicke model provides a paradigmatic example of quantum light-matter interactions~\cite{Dicke1954}. Its simplicity and capacity to encompass various collective phenomena, such as superradiant phase transitions~\cite{ Gross1982, Garraway2011, Kirton2018, Masson2022, Villaseor2024}, make the Dicke model a cornerstone in understanding light-matter interactions in the strong and ultra-strong coupling regimes~\cite{Diaz2019, Francisco2021, FriskKockum2019,le2020theoretical}. While the observation of the superradiant phase transition at equilibrium is hindered by renormalization terms~\cite{Rza1975,DeBernadis2018, Stefano2019, Stokes2019, Stokes2022}, here we focus on systems driven out of equilibrium. In this driven-dissipative case, several theoretical studies have predicted the emergence of phase transitions in the non-equilibrium steady states (NESS) as a function of the light-matter interaction strength~\cite{Bhaseen2012, Bastidas2012, Kirton2017, Kirton2018a, Masson2019, Reilly2024}. Further, dissipative phase transitions have been observed in a diverse range of experimental platforms, including ultra-cold atoms in optical cavities~\cite{Baumann2010, Hamner2014,  Klinder2015}, superconducting circuits~\cite{Brookes2021,Chen2023,FinkPRX17,beaulieu2023observation}, trapped ions~\cite{Cai2021}, nonlinear photonic or polaritonic modes~\cite{RodriguezPRL17, FinkNatPhys18, Zejian2022}, and other solid-state systems~\cite{Pallmann2023}. In the thermodynamic (TD) limit, the phenomenology of the Dicke model can be qualitatively understood with a semiclassical model, as both the NESS and its dynamical features can be accurately described by a mean-field (MF) approximation. Only around the critical point do higher-order correlations play a significant role~\cite{Emary2003, Nagy2010, Reiter2020}. 
\begin{figure}[t]
        \centering
        \includegraphics[width=1\linewidth]{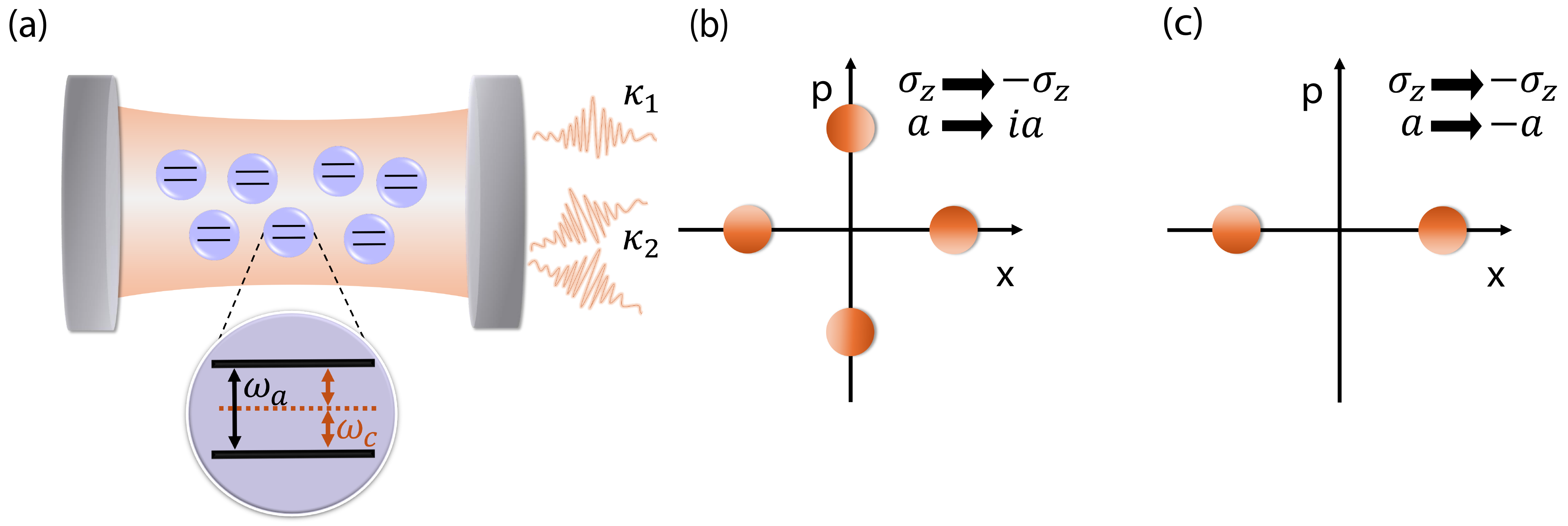}
        \caption{(a) Schematic diagram of an ensemble of identical two-level quantum emitters, with frequency $\omega_a$, coupled to a cavity with frequency $\omega_c$. The emitters are in resonance with two cavity photons, i.e. $\omega_a = 2\omega_c$. The decay rates, $\kappa_{1,(2)}$, represents one-(two) photon loss. Sketch of the phase space of the cavity mode highlighting (b) the $Z_4$-symmetry of the two-photon Dicke model, vs (c) the $Z_2$-symmetry of the standard Dicke model.}
        \label{fig: schematics}
\end{figure}

The development of the quantum engineering toolbox enables the realization of more complex light-matter interaction Hamiltonians. An interesting example is given by two-photon interactions, where the linear coupling is suppressed, and light and matter interact exclusively via the exchange of two excitation quanta. The implementation of these two-photon interactions is within reach of various platforms, such as trapped ions~\cite{Felicetti2015, Puebla2017, Casanova2018}, superconducting circuits~\cite{Felicetti2018, Felicetti2018-2, Gautier2022, Ayyash2024}, nanomechanical resonators~\cite{Wang2016}, mechanical modes coupled to spins~\cite{Sanchez2018}, and nonlinear photonic systems~\cite{Menotti2019, Zatti2023}.
Currently, growing research efforts are dedicated to the analysis of the mathematical~\cite{Travenec2012, Duan2016, Cong2019, Rico2020} and phenomenological~\cite{Cong2020, Zou2020, Shen2021, Piccione2022} properties of two-photon interactions.
Due to their strongly non-classical character, they have already been considered in various proposals for creating spin-squeezed states~\cite{Banerjee2022}, quantum state generation~\cite{Villas2019, Laha2024}, quantum batteries~\cite{Crescente2020, Delmonte2021}, the implementation of deterministic $CZ$ gates with propagating light~\cite{Alushi2023} and cavity photons~\cite{Tang2024}, cat-qubit stabilization~\cite{Gautier2022} and critical quantum sensing~\cite{Ying2022}. Although the two-photon Dicke model has been shown to manifest a superradiant phase transition in the Hamiltonian case~\cite{Garbe2017, Chen2018}, in the dissipative case the complexity of the model has so far limited the analysis to MF approaches~\cite{Garbe2020, Li2022, Li2024}.

In this work, we identify and analyze a dissipative phase transition in the two-photon quantum Dicke model. 
We first show that, beyond some critical coupling value, single-photon loss is insufficient to reach convergence, and the full quantum model is unstable. We then demonstrate that the inclusion of two-photon dissipation can fix the instability, and we identify a TD scaling for which a new stable phase appears beyond threshold. To explore the properties of these phases, we employ a combination of analytical and numerical techniques. Although MF predicts no stable solutions except for the normal phase, a second-order cumulant expansion~\cite{Kira2008, Reiter2020, Sanchez2020, Fowler-Wright2023} reveals three potential phases: one corresponding to the normal phase and two associated with superradiant states.  While mesoscopic systems are found to support two stable superradiant solutions, only one of them remains stable in the TD limit. Numerical simulations are performed using exact diagonalization (ED) for small system sizes and quantum trajectory (QT) simulations for larger systems, using QuTiP~\cite{QuTiP}. Our numerical results show excellent agreement with the phase boundaries obtained via the second-order cumulant analyses. These findings highlight the critical role of two-photon dissipation in enabling the stabilization of the rich dynamics of the two-photon Dicke model in open quantum systems.

\emph{The model --}  
We consider an ensemble of two-level quantum emitters coupled to a single-mode cavity, as depicted in Fig.~\ref{fig: schematics}(a). The dynamics can be described via the following Hamiltonian ($\hbar = 1$)
\begin{equation}~\label{eq: Hamiltonian-collective_spin}
    H = \omega_c a^\dagger a + \frac{\omega_a}{2} J_z + \frac{\lambda}{N} J_x(a^{\dagger 2} +  a^2) \, ,
\end{equation}
where $\omega_c$ and $\omega_a$ are the resonance frequencies of the cavity and the atom, respectively. $\lambda$ is the atom-light coupling strength, and $N$ is the total number of spins, also called the system size here. $J_{x,y,z}$ are collective spin operators that follow the usual commutation relation of $[J_x, J_y] = 2iJ_z$. We note that $1/N$ used in this work allows us to make the equations of motion $N$-independent, which can then be solved in the TD limit (cf.~the Supplemental Material (SM) for more information).
\begin{figure}[ht]
        \centering
        \includegraphics[width=1\linewidth]{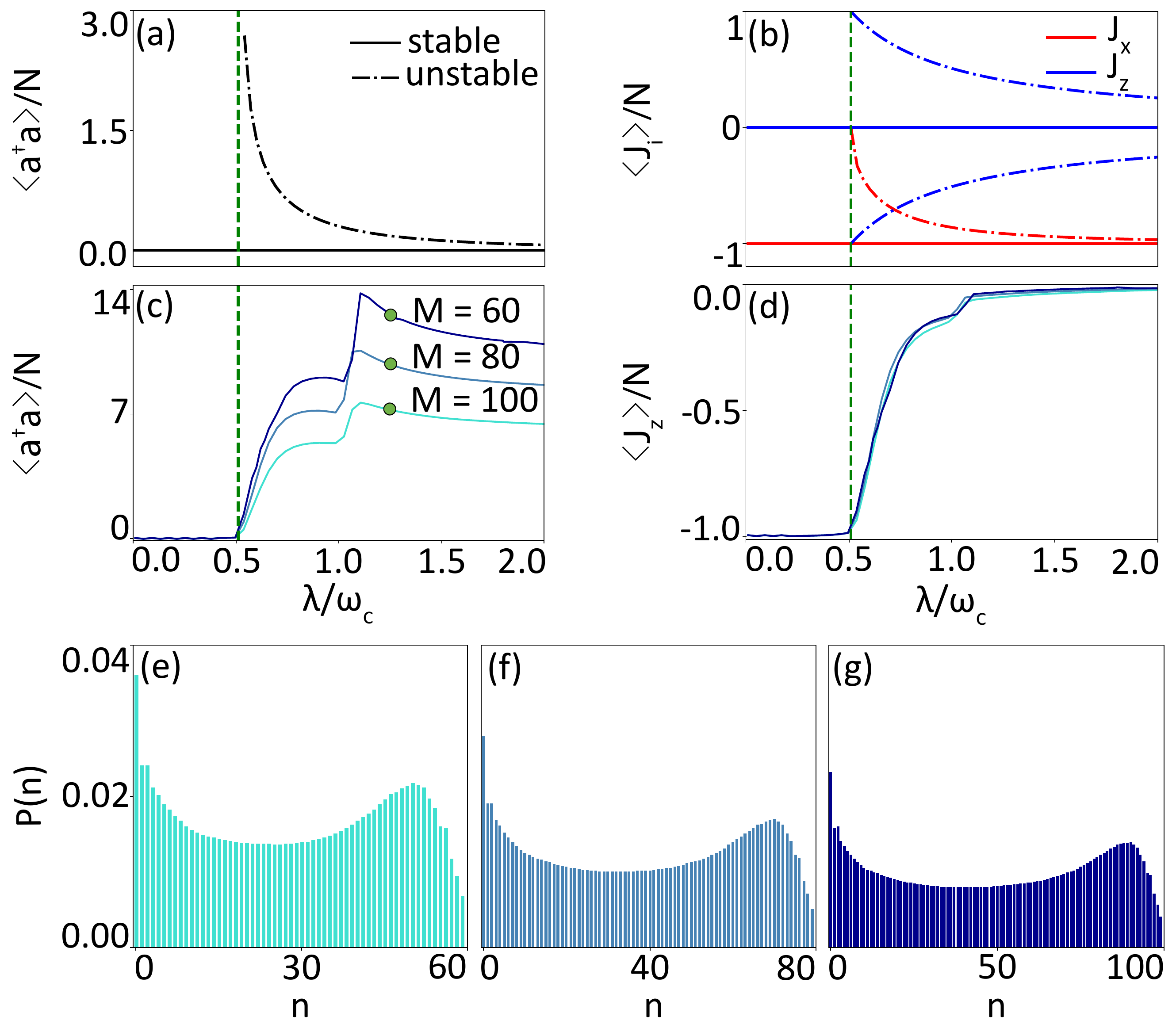}
        \caption{Instability of the open two-photon Dicke model with one-photon loss, only. The steady-state of the (a) average photons in the cavity and (b) the average collective spin, obtained from the analytic model. The solid (dashed) lines in each panel show the stable (unstable) solutions. (c) and (d) show the results of the full quantum mechanical calculations for $N = 4$ with various Fock state truncation numbers of $M = 60, 80, 100$. (e), (f), and (g) correspond to the photon probability distribution P(n) at the fixed coupling constants of $\lambda = 1.25$ and various truncation numbers of $M = 60, 80$, and 100, respectively. The non-vanishing occupation of higher number of states signals the instability of the model. In all cases, $\omega_a = \omega_c = 1$ and $\kappa_1 = 0.4$, and the green vertical dashed lines in panels (a)-(d) indicate the onset of the phase transition predicted by the analytical model.}
        \label{fig: 1-phloss}
\end{figure}   

As can be inferred from Eq.~\eqref{eq: Hamiltonian-collective_spin}, the Hamiltonian has a $Z_4$ symmetry (cf. Fig.~\ref{fig: schematics}(b)) which is distinct from the standard Dicke model with a $Z_2$ symmetry, as pictorially depicted in Fig.~\ref{fig: schematics}(c). In addition, the two-photon Dicke Hamiltonian is invariant under the generalized parity operator of $ \Pi$ = $e^{i\pi \left(a^\dagger a/2 + J_z\right)}$. This operator has four eigenvalues $\pm 1$ and $\pm i$, so the Hamiltonian can be divided into four block-diagonal subspaces. 

Moreover, $[H,J^2] = 0$, with $J^2=J_x^2 + J_y^2 + J_z^2 = N(N+1)$, which means that the total spin is a conserved quantity of the dynamics. There are only $N+1$ such states, since the spin ranges from -$N$ to $N$, so the size of the spin Hilbert space scales linearly with the system size. As detailed in the SM, the Heisenberg equations of motion for the photon or spin operators are $N$-independent via the following scaling
\begin{equation}~\label{eq: averaged}
    J_{x,y,z} \rightarrow \frac{J_{x,y,z}}{N} ~, ~ a \rightarrow \frac{a}{\sqrt{N}}\, .
\end{equation}
When the dynamics of the open quantum system are Markovian, the evolution of the density operator, $\rho$, is determined by the following Lindblad master equation
\begin{equation}~\label{eq: Liuvollian}
    \dot \rho = -i [H,\rho] + \sum_n \mathcal{D}[L_n]\rho \, ,
\end{equation}
where the dissipation related to the jump operator $L_n$ is read as  
\begin{equation}~\label{eq: jump operator1}
    \mathcal{D}[L_n]\rho = L_n \rho L_n^\dagger - \frac{1}{2} (L_n^\dagger L_n \rho + \rho L_n^\dagger L_n)\, .
\end{equation}
In this work, we include the loss mechanisms which correspond to photons leaking out of the cavity with
\begin{equation}
    L_1 = \sqrt{\kappa_1} a ~, ~ L_2 = \sqrt{\frac{\kappa_2}{N}} a^2\, ,
\end{equation}
where $\kappa_{1(2)}$ corresponds to the one (two)-photon loss rate and again the factor of $N$ is chosen to obtain $N$ independent MF equations in the TD limit. Since these jump operators do not commute with the generalized parity, $\Pi$, this $Z_4$ symmetry becomes a weak symmetry of the dynamics~\cite{Buca2023}. On the other hand, the total spin $J^2$ is still a good quantum number since the jumps are limited to only operators which affect the photons, rendering it a strong symmetry of the dynamics.

\emph{Dissipative phase transition --} We start our analysis by considering only single-photon loss$\kappa_2=0$ in Eq.~\eqref{eq: Liuvollian}. Figs.~\ref{fig: 1-phloss}(a) and (b), we show the NESS solution obtained by solving the equation of motion using MF theory. See the SM for details on how these equations and their solutions are obtained.
We see that below a critical value of $\lambda=\lambda_c$ (the vertical green dashed line in Figs.~\ref{fig: 1-phloss}(a)--(d)) the normal phase is the only solution to the equations and it is stable. Above this value, the MF equations admit a superradiant solution (black, blue, and red dashed lines in Figs.~\ref{fig: 1-phloss}(a) and (b)) but it is unstable and the normal state continues to be the only stable solution. 

We compare this result with what is obtained via ED for $N = 4$ in Fig.~\ref{fig: 1-phloss}(c) and (d). We see that above $\lambda_c$, the system finds a state with a finite population of the cavity mode; however, as the number of Fock states included in the exact simulation is increased, the system continues to populate higher and higher states without converging, implying that the dynamics is not stable. We see this non-convergence more clearly in Figs.~\ref{fig: 1-phloss}(e)--(g) where we show the diagonal elements of the photon density matrix, $P(n)$, corresponding to the occupation probability of the $n^\textrm{th}$ Fock state, for three different Fock space truncation numbers $M$.  This inability to reach a finite steady state is due to the fact that the included single-photon loss mechanism is unable to overcome the effective frequency renormalization due to two-photon couplings in the Hamiltonian~\cite{Travenec2012, Felicetti2015}. 
\begin{figure}[ht]
        \centering        
        \includegraphics[width=1\linewidth]{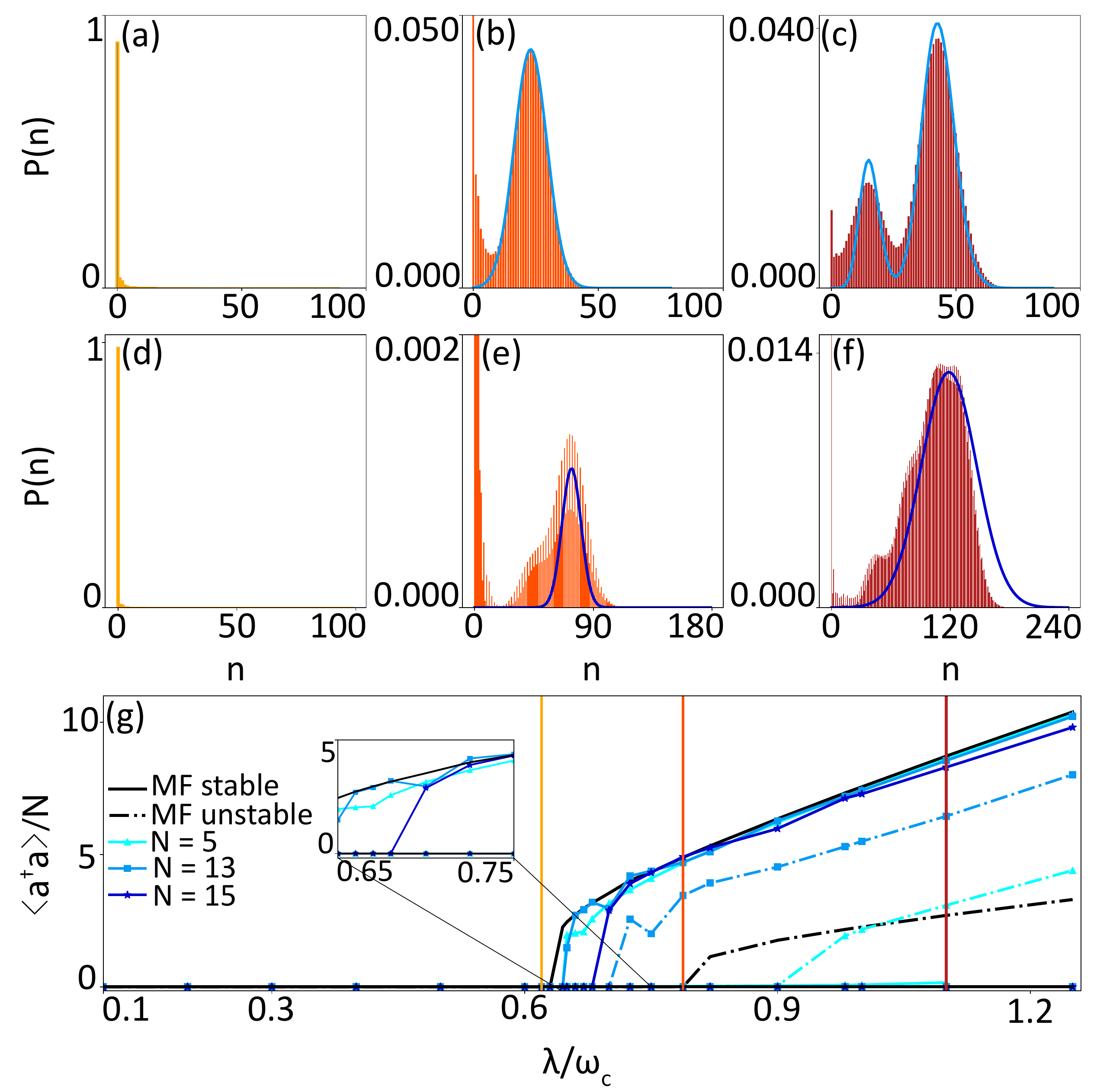}
        \caption{The stabilized open two-photon Dicke model with both one- and two-photon losses. (a), (b), and (c) show the occupation probability of the Fock state for small system size $N = 5$ spins at $\lambda = 0.62, 0.79$, and $1.1$ respectively. The light blue curve in (d) and (c) is the Gaussian fit used to extract the average values plotted in (g). (d), (e), and (f) shows the number state occupation probability for a larger system at $N = 15$ for the same coupling strength. The dark blue curve in (e) and (f) show the Gaussian fit to extract the average photons depicted in (g). (g) The solid (dashed) black lines show the stable (unstable) phases obtained from the analytical model. Different blue lines correspond to the average photon numbers for different system sizes extracted from the P(n) distributions depicted in (a)--(f). The inset shows the zoomed in region close to the phase transition point.}
        \label{fig: 2-phloss}
\end{figure}   

To counterbalance the effective pumping, we add a two-photon loss $L_2$ to the Liouvillian dynamics in Eq.~\eqref{eq: Liuvollian}. Figure~\ref{fig: 2-phloss}(a)--(c) presents the occupation probability of the number states for $N = 5$ for different values of the coupling $\lambda$, calculated via ED. Unlike the case of one-photon loss, here the occupation probability shows a clear convergence, i.e., there is no notable occupation of the higher Fock states. While for the weakly-interacting case at $\lambda = 0.62$, the vacuum is the prominently occupied state (cf. Fig.~\ref{fig: 2-phloss}(a)), for stronger coupling at $\lambda = 0.79$ an additional side lobe appears, which evolves into a bimodal feature at the very strong coupling limit at $\lambda = 1.1$ (cf. Fig.~\ref{fig: 2-phloss}(c)).

To study the scaling with system size and the emergence of any phase transition in the TD limit, we require the ability to find the NESS of larger systems. However, due to the unfavorable scaling of the size of the  Liouvillian as $((N+1)\times M)^2$, ED is not applicable. Instead, as detailed in the SM, we use a stochastic QT approach to calculate the density matrix. Figure~\ref{fig: 2-phloss}(d)--(f) shows the occupation probability of the number states for $N = 15$ at $\lambda = 0.62$, 0.79, and 1.1, respectively. For all three cases, the convergence and stability of the model are evident from the negligible population of the higher Fock states, which does not depend on the truncation number $M$. Similar to the $N = 5$ case, in weak coupling $n = 0$ is almost the only occupied state. An additional side lobe appears for moderate coupling, similar to the $N = 5$ case in Fig.~\ref{fig: 2-phloss}(e). However, unlike at small system sizes, for very strong coupling, as depicted in Fig.~\ref{fig: 2-phloss}(f), the Fock state distribution does not show the bimodal behavior. These exact results are indicative of both the stabilization of the model and the emergence of new solutions in addition to the normal phase.    

To better understand these results, we employ a second-order cumulant approximation to describe the dynamics of the averaged spin and photon quantities defined in Eq.~\eqref{eq: averaged} (see the SM for more information). The black lines in Fig.~\ref{fig: 2-phloss}(g) show the NESS of the average number of photons in the cavity vs.~the coupling strength, calculated from the analytical model. The solid and dashed-dotted lines delineate the stable and unstable solutions, determined via Bogoliubov analysis. While the normal state is always stable, in contrast to the case of one-photon loss, there is a critical coupling strength beyond which two superradiant solutions exist, of which only one of them is stable. 

To compare the results of this analytical model with the exact calculations of finite systems, we overlay the average photon numbers of $N = 5$, 13, and 15, calculated via ED or QT, in cyan, light blue, and dark blue, respectively. The vertical solid lines delimit the coupling strengths at which the Fock state occupations are depicted in Figs.~\ref{fig: 2-phloss}(a)--(f). The values of the superradiant solutions were obtained by using Gaussian fits to the corresponding Fock state distributions. The solid lines in Figs.~\ref{fig: 2-phloss}(a)--(f) show these Gaussian fits where the mean value of the fit corresponds to the average photon number for a particular value of $\lambda$ and gives  the dots in Fig.~\ref{fig: 2-phloss}(g). 

As can be seen, the analytical model describes the exact simulation results very well. It is also worthwhile to point out that the small discrepancy in the onset of transition for larger system sizes is a computational limitation. Since the probability of settling at the normal state is much higher in the vicinity of the phase-transition point, finding the superradiant solution requires many more quantum trajectories. In Fig.~\ref{fig: QMC convergence} of the SM, we show how by maximally randomizing the trajectories at $t = 0$ and increasing the number of trajectories, the agreement between the onset of the phase transition, in the analytical model and the numerical simulations, can be improved.

\begin{figure}[h]
        \centering
        \includegraphics[width=1\linewidth]{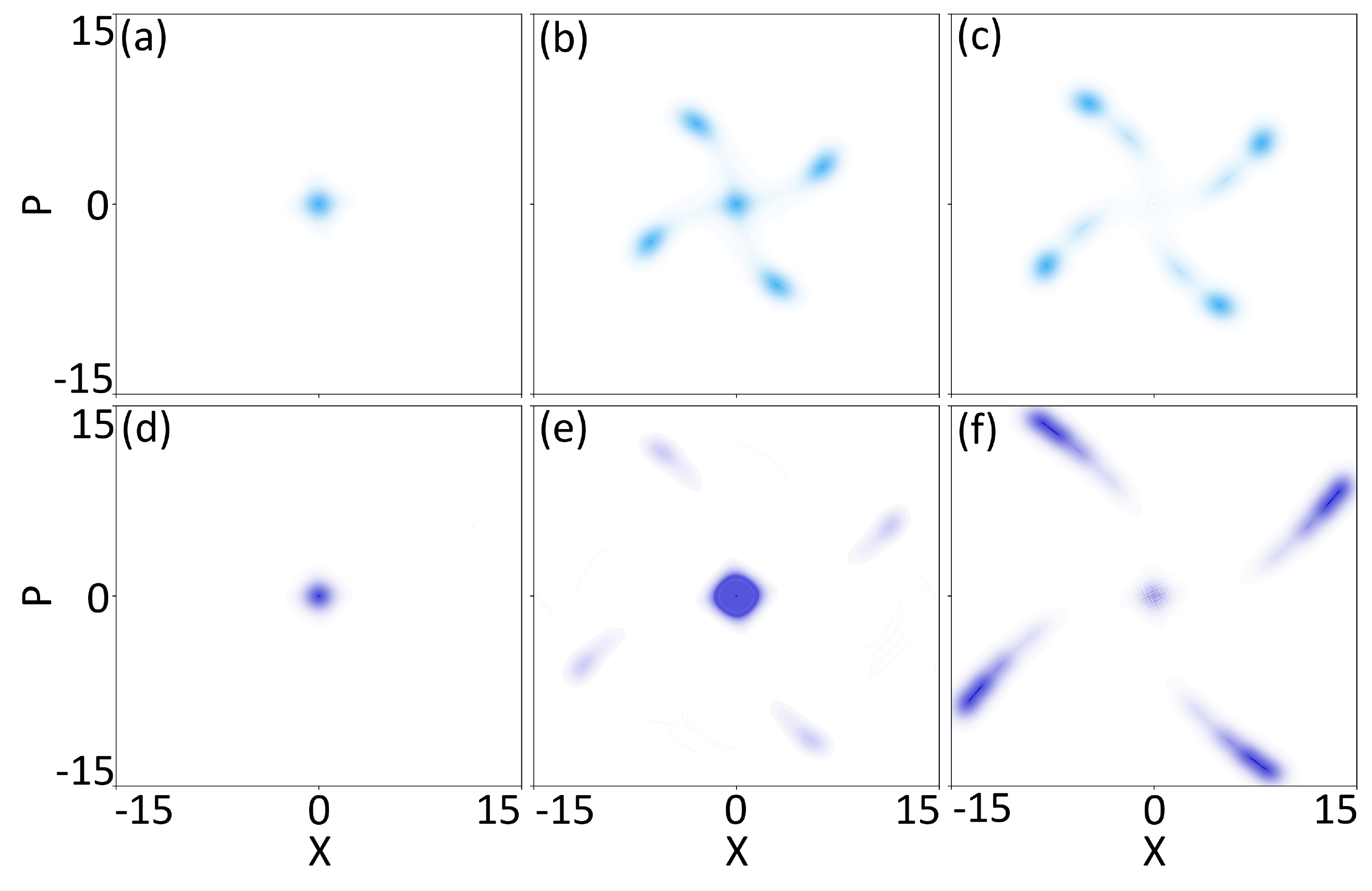}
        \caption{The photonic Wigner functions distribution of the stabilized two-photon Dicke model. Each panel here corresponds to panels with the same name in Fig.~\ref{fig: 2-phloss}. The colormap in panel (e) is saturated to highlight the side lobes corresponding to the superradiant solution.}
        \label{fig: Wigner}
\end{figure}

\emph{Wigner function --} 
To shed light on the type of phase transition and the symmetry properties of the solutions, in Fig.~\ref{fig: Wigner} we plot the Wigner distribution of the reduced photonic density matrix at various coupling strengths, $\lambda$, for two different system sizes $N$. Panels (a)--(f) present the Wigner functions corresponding to the cases in Fig.~\ref{fig: 2-phloss}(a)--(f). At small $\lambda$ before the phase transition, only the normal state exists and therefore there is no macroscopic occupation of the cavity. This can be clearly seen in Fig.~\ref{fig: Wigner}(a) and (d) for $N = 5$ and $N = 15$, respectively, where the Wigner function is very close to the vacuum. Figure~\ref{fig: Wigner}(b) and (e) show the Wigner function of the cavity photons at $\lambda = 0.79$, highlighted by the vertical red line in Fig.~\ref{fig: 2-phloss}(g), where in addition to the vacuum mode at the center there are four side lobes, corresponding to the other stable branches, both in small and large systems. The emergence of the four lobes in the superradiant phase shows the $Z_4$ symmetry of the Liouvillian. Furthermore, from the numerical evidence we have, it appears that the superradiant lobe is disjoint from the central vacuum state and coexists with it when it appears, this is behavior typical of a first order phase transition. Finally, at a strong coupling value of $\lambda = 1.1$, corresponding to the vertical red line in Fig.~\ref{fig: 2-phloss}(g), the Wigner functions of $N = 5$ and $N = 15$ are shown in Fig.~\ref{fig: Wigner}(c) and (f), respectively. For the large system, the Wigner function resembles the distribution of the weaker coupling in panel (e), i.e.~only the vacuum and the four side lobes are present. This agrees well with the fact that the normal state and the superradiant state are two stable TD phases. However, for smaller system sizes, the Wigner function depicted in Fig.~\ref{fig: Wigner}(c) has four additional, smaller, side lobes which correspond to the other superradiant state, only stable for finite-size systems. In all cases independent of system size and coupling strength, the numerically obtained Wigner function preserves the $Z_4$-symmetry; this is as expected for finite sized system, in the $N\to\infty$ limit these states become completely separated and so the symmetry can be spontaneously broken.

\emph{Conclusion -- }Here we presented exact solutions of a driven-dissipative two-photon Dicke model. Through detailed analyses, we showed that the inherent instability of the Hamiltonian cannot be stabilized via one-photon cavity loss, but adding even a small amount of the two-photon loss can stabilize the model for all system sizes. Further, through the second-order cumulant approximations, we derived an analytical model that describes the thermodynamic phases of the system as well as the finite-size system effects. Finally, calculations of the Wigner function highlighted the emergence of superradiant phases beyond the critical coupling and highlighted how the $Z_4$-symmetry of the model can be seen.
For future work, we propose more detailed studies to describe the nature of the phase transition and calculate its critical properties. In addition, one can study the robustness of different phases to spin decay and dephasing~\cite{Kirton2017, Lo2020, Rico2020, Roberts2020, Rota2019, Shammah2018, Wang2019, Xu2023}. Beyond the fundamental interest, these unconventional properties might find direct application in critical quantum sensing~\cite{ding2022enhanced, beaulieu2024} and in the encoding of cat qubits~\cite{Gravina23}.

\section*{Acknowledgment}
\noindent
Research supported as part of QuPIDC, an Energy Frontier Research Center, funded by the US Department of Energy (DOE), Office of Science, Basic Energy Sciences (BES), under award number DE-SC0025620 (AJS and HA).
In addition, part of the research was supported by the US Department of Energy (DOE), the Office of Basic Energy Sciences (BES), and the Division of Materials Sciences and Engineering under award number DE-SC0025554 (AJS and HA). This research was supported in part by grant number NSF PHY-2309135 to the Kavli Institute for Theoretical Physics (HA).\\

\clearpage
\newpage
\bibliography{ref}

\clearpage
\newpage
\setcounter{equation}{0}

\section{Equation of motion for one-photon loss}~\label{app: EoM}
For the two-photon Dicke model subject to the one-photon loss, the Heisenberg equations of motion for the cavity field and collective spin operators have the following form
\begin{align}
    %
    %
   \langle \dot {a^2} \rangle & = - \left(\kappa_1 + 2i\omega_c \right) \langle a^2 \rangle - \frac{4i\lambda}{N} \langle J_x n \rangle \,, \\
    {\langle \dot {a^{\dagger^2}}\rangle} & = - \left(\kappa_1 - 2i\omega_c \right) \langle a^{\dagger^2} \rangle + \frac{4i\lambda}{N} \langle J_x n \rangle \,, \\
   \langle \dot n \rangle & = - \kappa_1 \langle n \rangle + \frac{2i\lambda}{N} \left(\langle J_x a^2 \rangle - \langle J_x a^{\dagger^2} \rangle \right) \,, \\
     \langle \dot {J_x} \rangle & = - \frac{2\omega_a}{N} \langle J_y \rangle \,, \\
    \langle \dot J_y \rangle & = \frac{2\omega_a}{N} \langle J_x \rangle - \frac{2\lambda}{N} \left(\langle J_z a^2 \rangle + \langle J_z a^{\dagger^2} \rangle \right)\,, \\
     \langle \dot J_z \rangle & =  \frac{2\lambda}{N} \left(\langle J_y a^2 \rangle + \langle {J_y} a^{\dagger^2} \rangle \right)\, ,
\end{align}
where $n = a^\dagger a$ is the photon number operator.

Throughout this work, we used the MF approximation to separate spin-photon correlations as 
\begin{equation}
    \langle \hat{O}_\textrm{spin} \hat{O}_\textrm{photon} \rangle = \langle \hat{O}_\textrm{spin} \rangle \langle \hat{O}_\textrm{photon} \rangle\, .
\end{equation}
As mentioned in the main text, one can arrive at a set of $N$-independent equations through the scaling introduced in Eq.~\eqref{eq: averaged}. The scaled and spin-photon separated equations in the TD limit read as 
\begin{align}
    \partial_t \braket{X} & = -\kappa_1 \braket{X} - 2i \omega_c \braket{Y} \,, \\
    \partial_t \braket{Y} & = -\kappa_1 \braket{Y} - 2i\omega_c \braket{X} - i 8 \lambda \braket{J_x} \braket{n} \,, \\
    \partial_t \braket{n} & = -\kappa_1 \braket{n}  + i 2 \lambda \braket{J_x} \braket{Y} \,, \\ 
    \partial_t \braket{J_x} & = -2 \omega_a \braket{J_y}  \,, \\
    \partial_t \braket{J_y} & = 2 \omega_a \braket{J_x} - 2 \lambda \braket{J_z} \braket{X} \,, \\
    \partial_t \braket{J_z}  & = 2 \lambda \braket{J_y} \braket{X} \,,
\end{align}
where $X = a^2 + a^{\dagger^2}$ and $Y = a^2 - a^{\dagger^2}$.

By solving for a steady state (ss), two different phases can be identified, where the phase transition occurs at the critical coupling $\lambda_c$ as 
\begin{equation}
    \lambda_c = \frac{1}{4} \sqrt{\kappa^2 + 4\omega_c^2}\, .
\end{equation}

1) The normal phase for $\lambda \le \lambda_c$, where all spins are in the ground state and there is no macroscopic occupation of the cavity.
$\langle X \rangle_{ss} = \langle Y \rangle_{ss} = \langle n \rangle_{ss} = \langle J_x \rangle_{ss} = \langle J_y \rangle_{ss} = 0 $ and $\langle J_z \rangle_{ss} = -1$.

2) The superradiant phase for $\lambda \ge \lambda_c$, where the spins are polarized, and there is a non-zero cavity occupation.
$\langle X \rangle_{ss} = \langle Y \rangle_{ss} = \langle n \rangle_{ss} = \langle J_x \rangle_{ss} = \langle J_z \rangle_{ss} \neq 0 $ and $\langle J_y \rangle_{ss} = 0$. 


Since there is no spin-related jump operators, the total spin is a conserved quantity, i.e. the strong symmetry, as 
\begin{equation}
    \langle J_x \rangle_{ss}^2 + \langle J_y \rangle_{ss}^2 + \langle J_z \rangle_{ss}^2 = 1
\end{equation}
After the PT, $J_y = 0$, therefore
\begin{equation}
    \langle J_x \rangle_{ss}^2 + \langle J_z \rangle_{ss}^2 = 1\, .
\end{equation}
The explicit solutions for the SP read as follows
\begin{align}
    \langle J_x \rangle_{ss} & = \pm \frac{\lambda_c}{\lambda}  \,, \\
    \langle J_y \rangle_{ss} & = 0 \,, \\
    \langle J_z \rangle_{ss} & = -\sqrt{1 - \frac{\lambda_c^2}{\lambda^2}} \,, \\ 
    \langle n \rangle_{ss} & = -\frac{\omega_a}{\omega_c} \frac{\langle J_x \rangle_{ss}^2}{\langle J_z \rangle_{ss}}  \,, \\
    \langle X \rangle_{ss} & = \frac{\omega_a}{\lambda}\frac{\langle J_x \rangle_{ss}}{\langle J_z \rangle_{ss}} \,, \\
    \langle Y \rangle_{ss}  & = - \frac{\kappa_1}{2i \omega_c} \langle X \rangle_{ss} \,,
\end{align}    

\subsection{Linear Stability Analysis}~\label{app: MF stability analysis}
To check the stability of the steady-state solutions obtained in the previous section, we perform a linear stability analysis. 

Let $\eta(t) = x(t) - x^*$ be a small perturbation around $x^*$, a fixed point of the EoM. Inserting this ansatz back into the scaled EoM and linearizing them around the fixed point by retaining only linear terms in perturbation, i.e. O($\eta$) we arrive at 
\begin{equation}
    \dot\eta(t) = \mathcal{B}(x^*) \eta \, .
\end{equation}
where $\mathcal{B}(x^*)$, aka the Bogoliubov matrix, is formally the small perturbation matrix around the fixed point $x^*$. If the spectrum of this matrix is always within the left-half of the complex plane, then the solution is stable, and otherwise it is unstable. 

For the normal state, the Bogoliubov matrix $\mathcal{B}$ is read as
\begin{equation}
    \mathcal{B} = 
\begin{bmatrix}
    -\kappa_1 & -2i \omega_c & 0 & 0 & 0 & 0 \\
    -2i \omega_c & - \kappa_1 & 0 & 0 & 0 & 0 \\
    0 & 0 & -\kappa_1 & 0 & 0 & 0 \\
    0 & 0 & 0 & 0 & -2\omega_a & 0 \\
    2\lambda & 0 & 0 & 2\omega_a & 0 & 0 \\
    0 & 0 & 0 & 0 & 0 & 0 \\
\end{bmatrix}
\end{equation}
with the following eigenvalues
\begin{equation}
 -\kappa_1 \pm 2i\omega_c,  -\kappa_1, \pm 2i \omega_a, 0\, .
\end{equation}
which implies that the normal state is always stable.

\begin{figure}
    \centering
    \includegraphics[width=1\linewidth]{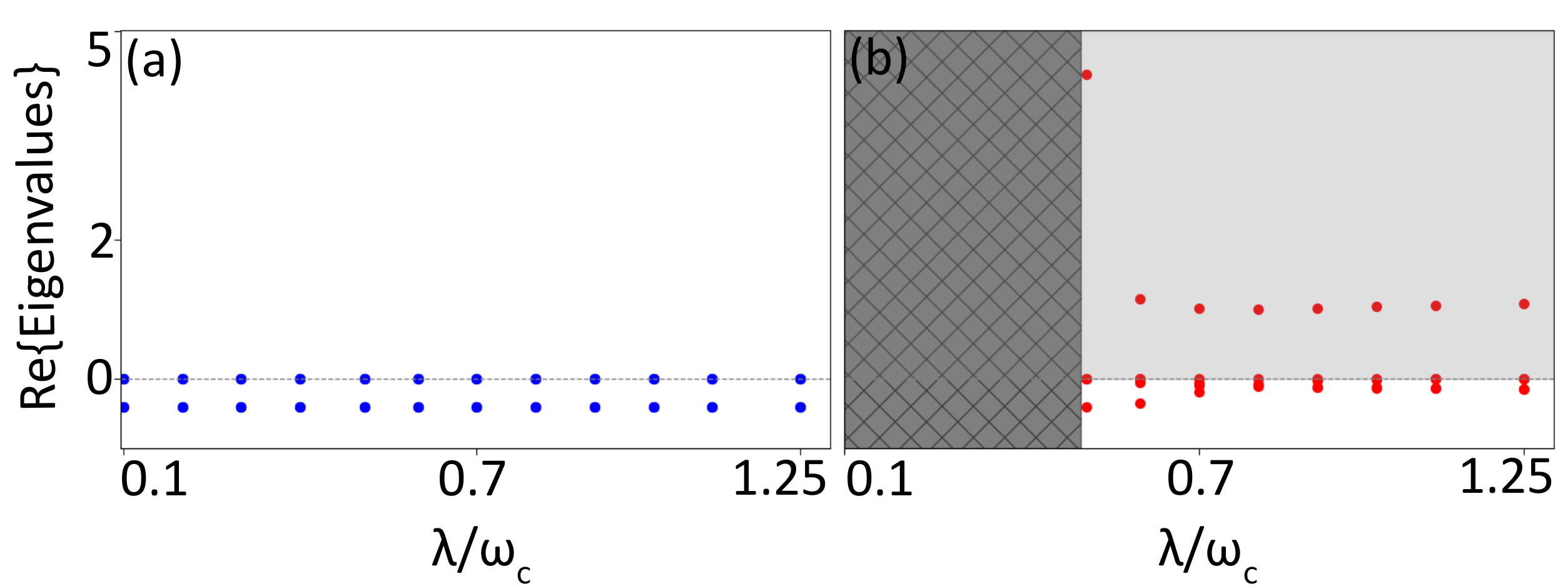}
    \caption{Real part of the Bougolibov matrix spectrum for the (a) normal state, showing stable solution for all values of $\lambda$, and (b) superradiant state showing unstable solution as some real parts of the eigenvalues are greater than zero, indicated by the shaded gray region. The hashed regions in (b) for $\lambda \leq 0.51$ delimit the absence of the state solution in that coupling range. }
    \label{fig: 1phMF-eigh}
\end{figure}

For the superradiant state the stability matrix reads as
\begin{widetext}
\begin{equation}
    \mathcal{B} = 
\begin{bmatrix}
    -\kappa_1 & -2i \omega_c & 0 & 0 & 0 & 0 \\
    -2i \omega_c & - \kappa_1 & -8i\lambda \langle J_x \rangle_{ss} & -8i\lambda \langle n \rangle_{ss} & 0 & 0 \\
    0 & 2i\lambda \langle J_x \rangle_{ss} & -\kappa_1 & 2i\lambda \langle Y \rangle_{ss} & 0 & 0 \\
    0 & 0 & 0 & 0 & -2\omega_a & 0 \\
    -2\lambda \langle J_z \rangle_{ss} & 0 & 0 & 2\omega_a & 0 & -2\lambda \langle X \rangle_{ss} \\
    2\lambda \langle J_y \rangle_{ss} & 0 & 0 & 0 & 2\lambda \langle X \rangle_{ss} & 0 \\
\end{bmatrix}
\end{equation}
\end{widetext}
The eigenvalues for the superradiant phase can be numerically determined for different steady-state values. As shown in Fig.~\ref{fig: 1phMF-eigh} there are always eigenvalues with positive real parts, which signal the instability, as highlighted by the gray region in panel (b).
\section{Equations of motion for one- and two-photon losses}~\label{app: EoM12}
The inclusion of the two-photon loss modifies the equations of motion as follows
\begin{widetext}
\begin{align}
    \partial_t \langle a \rangle & = - \left(\frac{\kappa_1}{2} + i \omega_c \right) \langle a \rangle - 2i\frac{\lambda}{N} \langle J_x \rangle \langle a^\dagger \rangle -\frac{\kappa_2}{N} \langle a^\dagger a^2 \rangle\,, \\
    \partial_t \langle a^\dagger \rangle & = - \left(\frac{\kappa_1}{2} - i \omega_c \right) \langle a^\dagger \rangle + 2i\frac{\lambda}{N} \langle J_x \rangle \langle a \rangle -\frac{\kappa_2}{N} \langle a a^{\dagger2} \rangle\,, \\
    \partial_t \langle a^2 \rangle & = - \left(\kappa_1 + 2i\omega_c \right)\langle a^2 \rangle - 4i\frac{\lambda}{N} \langle J_x \rangle \langle n \rangle -2\frac{\kappa_2}{N} (\langle a^2 \rangle + 2 \langle a^\dagger a^3\rangle)\,, \\
    \partial_t \langle a^{\dagger2} \rangle & = - \left(\kappa_1 - 2i\omega_c \right)\langle a^{\dagger2} \rangle + 4i\frac{\lambda}{N} \langle J_x \rangle \langle n \rangle -2\frac{\kappa_2}{N} (\langle a^{\dagger2} \rangle + 2 \langle a a^{\dagger3}\rangle)\,, \\
    \partial_t \langle n \rangle & = -\kappa_1 n + 2i\frac{\lambda}{N} \langle J_x \rangle (\langle a^2 \rangle - \langle a^{\dagger2} \rangle) -\frac{\kappa_2}{N}\langle a^{\dagger^2} a^2 \rangle \,, \\ 
    \partial_t \langle J_x \rangle & = -2 \omega_a \langle J_y \rangle \,, \\
    \partial_t \langle J_y \rangle & = 2 \omega_a \langle J_x \rangle - 2\frac{\lambda}{N}\langle J_z \rangle (\langle a^2 \rangle + \langle a^{\dagger2} \rangle)  \,, \\
    \partial_t \langle J_z \rangle & = 2 \frac{\lambda}{N} \langle J_y \rangle(\langle a^2 \rangle + \langle a^{\dagger2} \rangle)  \,.
\end{align}
\end{widetext}
Unlike the one-photon loss case where photon correlators appear to the quadratic order at most, here one gets higher-order photon correlations. We employ the second-order cumulant approximation to express these higher-order correlations in terms of the first- and second-order ones and obtain the following equations
\begin{widetext}
\begin{align}
    \partial_t \langle a \rangle & = - \left(\frac{\kappa_1}{2} + i \omega_c \right) \langle a \rangle - 2i\frac{\lambda}{N} \langle J_x \rangle \langle a^\dagger \rangle -\frac{\kappa_2}{N} (2\langle n \rangle \langle a \rangle + \langle a^2 \rangle \langle a^\dagger \rangle - 2 \langle a \rangle^2 \langle a^\dagger \rangle)\,, \\
     \partial_t \langle a^\dagger \rangle & = - \left(\frac{\kappa_1}{2} - i \omega_c \right) \langle a^\dagger \rangle + 2i\frac{\lambda}{N} \langle J_x \rangle \langle a \rangle -\frac{\kappa_2}{N} (2\langle n \rangle \langle a^\dagger \rangle + \langle a^{\dagger2} \rangle \langle a \rangle - 2 \langle a^\dagger \rangle^2 \langle a \rangle)\,, \\
    \partial_t \langle a^2 \rangle & = - (\kappa_1 + 2i\omega_c)\langle a^2 \rangle - 4i\frac{\lambda}{N} \langle J_x \rangle \langle n \rangle -2\frac{\kappa_2}{N} (3\langle n \rangle \langle a^2 \rangle -2\langle a^\dagger \rangle \langle a \rangle^3)\,, \\
    \partial_t \langle a^{\dagger2} \rangle & = - (\kappa_1 - 2i\omega_c)\langle a^{\dagger2} \rangle + 4i\frac{\lambda}{N} \langle J_x \rangle \langle n \rangle -2\frac{\kappa_2}{N} (3\langle n \rangle \langle a^{\dagger2} \rangle -2\langle a \rangle \langle a^\dagger \rangle^3)\,, \\
    \partial_t \langle n \rangle & = -\kappa_1 n + 2i\frac{\lambda}{N} \langle J_x \rangle (\langle a^2 \rangle - \langle a^{\dagger2} \rangle) -\frac{\kappa_2}{N}(4 \langle n \rangle^2 -4 \langle a^\dagger \rangle^2 \langle a \rangle^2  + 2\langle a^{\dagger2} \rangle \langle a^2 \rangle) \,, \\ 
    \partial_t \langle J_x \rangle & = -2 \omega_a \langle J_y \rangle \,, \\
    \partial_t \langle J_y \rangle & = 2 \omega_a \langle J_x \rangle - 2\frac{\lambda}{N}\langle J_z \rangle (\langle a^2 \rangle + \langle a^{\dagger2} \rangle)  \,, \\
    \partial_t \langle J_z \rangle & = 2 \frac{\lambda}{N} \langle J_y \rangle(\langle a^2 \rangle + \langle a^{\dagger2} \rangle)  \,.
\end{align}
\end{widetext}
After appropriately scaling the operators as in Eq.~\eqref{eq: averaged} we arrive at the following equations in the TD limit
\begin{widetext}
\begin{align}
    \partial_t \langle a \rangle & = - \left(\frac{\kappa_1}{2} + i \omega_c \right) \langle a \rangle - 2i\lambda \langle J_x \rangle \langle a^\dagger \rangle - \kappa_2 (2\langle n \rangle \langle a \rangle + \langle a^2 \rangle \langle a^\dagger \rangle - 2 \langle a \rangle^2 \langle a^\dagger \rangle)\, , \\
    \partial_t \langle a^\dagger \rangle & = - \left(\frac{\kappa_1}{2} - i \omega_c \right) \langle a^\dagger \rangle + 2i\lambda \langle J_x \rangle \langle a \rangle - \kappa_2 (2\langle n \rangle \langle a^\dagger \rangle + \langle a^{\dagger2} \rangle \langle a \rangle - 2 \langle a^\dagger \rangle^2 \langle a \rangle)\, , \\
    \partial_t \langle a^2 \rangle & = - \left(\kappa_1 + 2i\omega_c \right) \langle a^2 \rangle - 4i\lambda \langle J_x \rangle \langle n \rangle -2\kappa_2 ( 3\langle n \rangle \langle a^2 \rangle -2\langle a^\dagger \rangle \langle a \rangle^3)\, , \\
    \partial_t \langle a^{\dagger2} \rangle & = - \left(\kappa_1 - 2i\omega_c \right) \langle a^{\dagger2} \rangle + 4i\lambda \langle J_x \rangle \langle n \rangle -2\kappa_2 ( 3\langle n \rangle \langle a^{\dagger2} \rangle -2\langle a \rangle \langle a^\dagger \rangle^3)\, , \\    
    \partial_t \langle n \rangle & = 2i\lambda \langle J_x \rangle (\langle a^2 \rangle - \langle a^{\dagger2} \rangle) - \kappa_1 \langle n \rangle - \kappa_2(4 \langle n \rangle^2 -4 \langle a^\dagger \rangle^2 \langle a \rangle^2  + 2\langle a^{\dagger2} \rangle \langle a^2 \rangle) \, , \\
    \partial_t \langle J_x \rangle & = -2 \omega_a \langle J_y \rangle\, , \\
    \partial_t \langle J_y \rangle & = 2 \omega_a \langle J_x \rangle - 2\lambda \langle J_z \rangle (\langle a^2 \rangle + \langle a^{\dagger2} \rangle) \, , \\
    \partial_t \langle J_z \rangle & = 2\lambda \langle J_y \rangle(\langle a^2 \rangle + \langle a^{\dagger2} \rangle)\, .
\end{align} 
\end{widetext}
These equations can be solved analytically to determine the steady states and identify various normal and superradiant states, as discussed for the one-photon loss in the previous section.
\subsection{Linear stability analysis}
\begin{figure}
    \centering
    \includegraphics[width=1\linewidth]{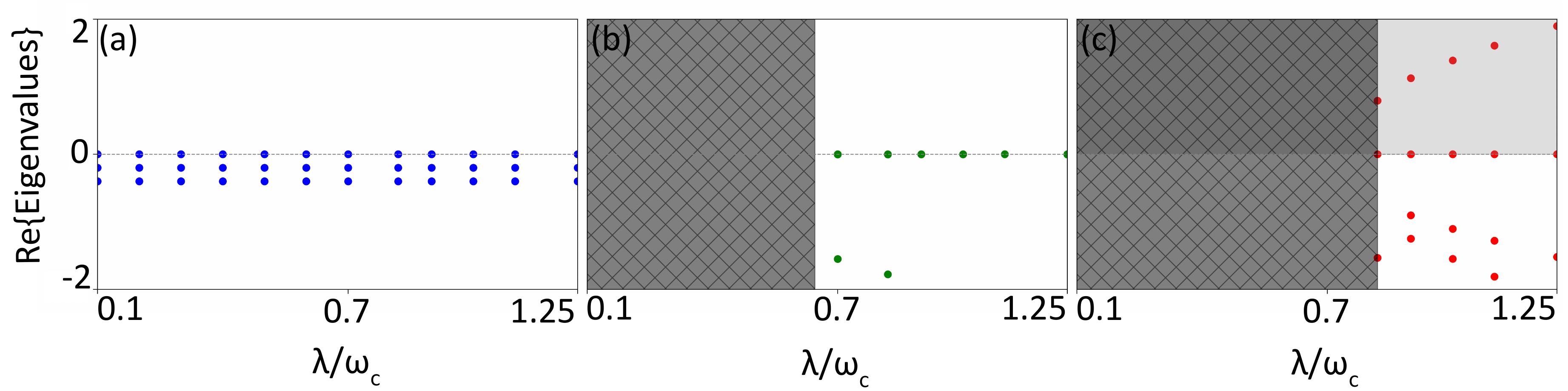}
    \caption{Real part of the Bougolibov matrix spectrum vs. $\lambda$, for the (a) normal state, showing the stablility of the solution for all interaction strengths, (b) larger superradiant state showing stability of the solution as all real parts of the eigenvalues are less than or equal to zero, and (c) smaller superradiant state which is unstable due to positive values highlighted by the gray shaded region. The hashed regions in (b) for $\lambda \leq 0.645$ and in (c) for $\lambda \leq 0.82$ delimit the absence of the solution in that coupling range.}
    \label{fig: 1and2phMF-eigh}
\end{figure}

Following the procedure detailed for the one-photon case, we obtain the Bogoliubov matrix as
\begin{widetext}
\begin{equation}
    \mathcal{B} = 
\begin{bmatrix}
    -\left(\frac{\kappa_1}{2} +i\omega_c \right) & 0 & 0 & 0 & 0 & 0 &0 & 0\\
    0 & -\left(\frac{\kappa_1}{2} -i\omega_c \right) & 0 & 0 & 0 & 0 &0 &0 \\
    0 & 0 & -(\kappa_1 + 2i\omega_c) & 0 & 0 & 0 & 0 & 0\\
    0 & 0 & 0 & -(\kappa_1 - 2i\omega_c) & 0 & 0 & 0 & 0\\
    0 & 0 & 0 & 0 & -\kappa_1  & 0 & 0 & 0 \\
    0 & 0 & 0 & 0 & 0 & 0 & -2\omega_a & 0 \\
    0 & 0 & 2\lambda & 2\lambda & 0 & 2\omega_a & 0 & 0 \\
    0 & 0 & 0 & 0 & 0 & 0 & 0 & 0\\
\end{bmatrix}\, .
\end{equation}
\end{widetext}
This has the same form as for the one-photon loss, and hence the stability of the normal state is guaranteed by its spectrum in the left half of the complex plane, for all values of $\lambda$, as can be seen in Fig.~\ref{fig: 1and2phMF-eigh}(a).

For the superradiant states, however, we have
\begin{widetext}
\begin{equation}
    \mathcal{B} = 
\scalebox{0.5}{$
\begin{bmatrix}
    -(\frac{\kappa_1}{2} +i\omega_c) - 2\kappa_2 \langle n \rangle_{ss} + 4\kappa_2 \langle a \rangle_{ss} \langle a^\dagger \rangle_{ss}& -2i\lambda \langle J_x \rangle_{ss} - \kappa_2 \langle a^2 \rangle_{ss} + 2\kappa_2 \langle a \rangle_{ss}^2 & - \kappa_2 \langle a^\dagger \rangle_{ss} & 0 & - 2\kappa_2 \langle a \rangle_{ss} & -2i\lambda \langle a^\dagger \rangle_{ss} & 0 & 0 \\
    2i\lambda \langle J_x \rangle_{ss} - \kappa_2 \langle a^{\dagger2} \rangle_{ss} + 2\kappa_2 \langle a^\dagger \rangle_{ss}^2 & -(\frac{\kappa_1}{2} - i\omega_c) - 2\kappa_2 \langle n \rangle_{ss} + 4\kappa_2 \langle a^\dagger \rangle_{ss} \langle a \rangle_{ss} & 0 & -\kappa_2 \langle a \rangle_{ss} &- 2\kappa_2 \langle a^\dagger \rangle_{ss} & 2i\lambda \langle a \rangle_{ss} & 0 & 0 \\
    12\kappa_2 \langle a^\dagger \rangle \langle a \rangle^2_{ss} & 4\kappa_2 \langle a \rangle_{ss}^3 & -(\kappa_1 + 2i\omega_c) - 6\kappa_2 \langle n \rangle_{ss} & 0 & -4i\lambda \langle J_x \rangle_{ss} - 6\kappa_2 \langle a^2 \rangle_{ss} & -4i\lambda \langle n \rangle_{ss} & 0 & 0 \\
    4\kappa_2 \langle a^\dagger \rangle_{ss}^3 &  12\kappa_2 \langle a \rangle \langle a^\dagger \rangle^2_{ss} & 0 & -(\kappa_1 - 2i\omega_c) - 6\kappa_2 \langle n \rangle_{ss} & 4i\lambda \langle J_x \rangle_{ss} - 6\kappa_2 \langle a^{\dagger2} \rangle_{ss} & 4i\lambda \langle n \rangle_{ss} & 0 & 0 \\
    8\kappa_2 \langle a 
    \rangle_{ss} \langle a^\dagger \rangle^2_{ss} & 8\kappa_2 \langle a^\dagger \rangle_{ss} \langle a \rangle^2_{ss} & 2i\lambda \langle J_x \rangle_{ss} -2\kappa_2 \langle a^{\dagger2} \rangle_{ss}& -2i\lambda \langle J_x \rangle_{ss} -2\kappa_2 \langle a^2 \rangle_{ss} & -\kappa_1 -8\kappa_2 \langle n \rangle_{ss} & 2i\lambda (\langle a^2 \rangle_{ss} - \langle a^{\dagger2} \rangle_{ss}) & 0 & 0 \\
    0 & 0 & 0 & 0 & 0 & 0 & -2\omega_a & 0 \\
    0 & 0 & -2\lambda \langle J_z \rangle_{ss} & -2\lambda \langle J_z \rangle_{ss} & 0 & 2\omega_a & 0 & -2\lambda (\langle a^2 \rangle_{ss} + \langle a^{\dagger2} \rangle_{ss}) \\
    0 & 0 & 2\lambda \langle J_y \rangle_{ss} & 2\lambda \langle J_y \rangle_{ss} & 0 & 0 & 2\lambda (\langle a^2 \rangle_{ss} + \langle a^{\dagger2} \rangle_{ss}) & 0 \\
\end{bmatrix}\, .
$}
\end{equation}
\end{widetext}
For the two superradiant states obtained from the numerical solution of the equations of motion, we find that the upper branch in Fig.~\ref{fig: 2-phloss}(g) is stable for all values of $\lambda$, while the second branch is unstable always due to the existence of eigenvalues in the right half-plane, as illustrated in Fig.~\ref{fig: 1and2phMF-eigh}(b) and (c), respectively.
\section{Quantum trajectory}~\label{app: QMC}
As mentioned in the main text, for larger $N$, obtaining the full solution by ED is not feasible, so we used the stochastic QT approach instead. To randomize enough the state vector at $t = 0$ to capture subtle features especially close to the phase transition point, we randomly draw a state from a density matrix at infinite temperature, i.e., all possible states have equal probability of being selected.

If $\ket{\psi_i}$ is the $i^\textrm{th}$ trajectory, one can determine the density matrix $\hat{\rho}$ as 
\begin{equation}
    \hat{\rho} = \sum_i^{N_T} \frac{\ket{\psi_i} \bra{\psi_i}}{N_T}\, ,
\end{equation}
where $N_T$ is the number of trajectories. 

To calculate photonic properties such as the probability of Fock state occupation as well as the Wigner function, we calculated the reduced photonic density matrix as $\hat{\rho}_\textrm{ph} = \textrm{Tr}_s(\hat{\rho})$, where $\textrm{Tr}_s$ represents the partial trace operator over all spin degrees of freedom. 

\begin{figure}[h]
        \centering
        \includegraphics[width=0.85\linewidth]{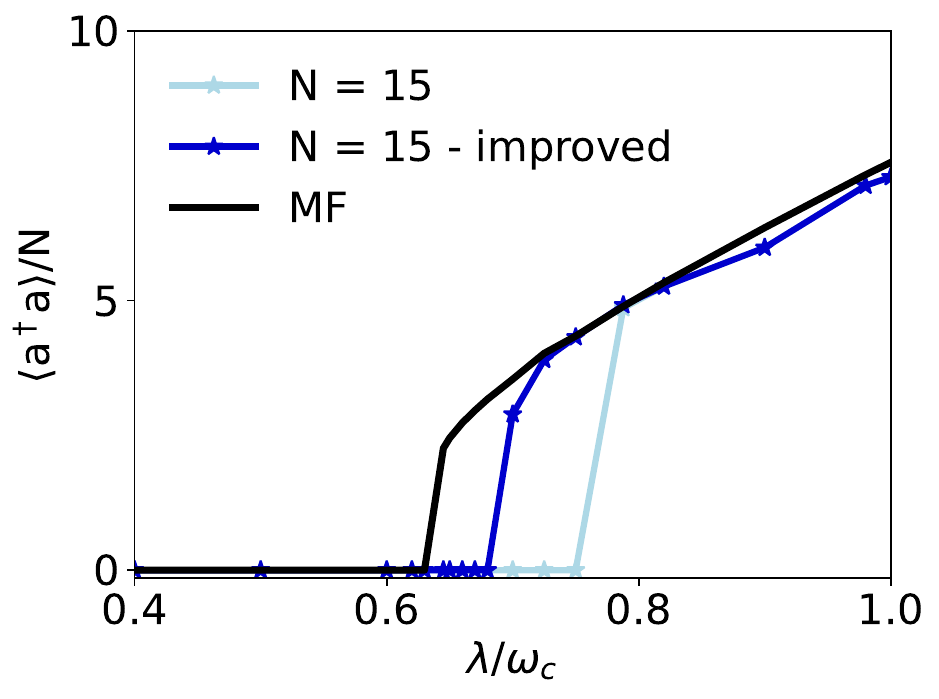}
        \caption{Convergence of the quantum trajectory approach for $N = 15$ for $N_T$ = 500 (light blue) and $N_T$ = 3000 (dark blue). The solid black line shows the analytic results in the thermodynamic limit for comparison.}
        \label{fig: QMC convergence}
\end{figure}

The convergence of the results was confirmed by not having any change after increasing the simulation time and the number of trajectories $N_T$. Figure~\ref{fig: QMC convergence} shows the variation of the average photon number vs. $\lambda$ for $N = 15$ when the number of trajectories increased from 500 to 3000, denoted by the light blue and dark blue curves, respectively. The analytical results are shown in black to highlight the good agreement between QT and the analytic results as the number of trajectories increases. We attribute the remained discrepancy between the dark blue and black lines close to the phase transition to computational limitations.
\section{Symmetry of the two-photon loss only}~\label{app: Liouv}
If in Eq.~\eqref{eq: Liuvollian}, only $L_2$, i.e. a two-photon loss mechanism exists, the photon exchanges always happen in pairs. Therefore, the even and odd number states cannot get coupled to each other throughout the dynamics. It means that the Liouvillian superoperator has a degenerate kernel, and hence there are two distinct NESSes with even-only and odd-only Fock states. 

\begin{figure}[h]
        \centering        
        \includegraphics[width=1\linewidth]{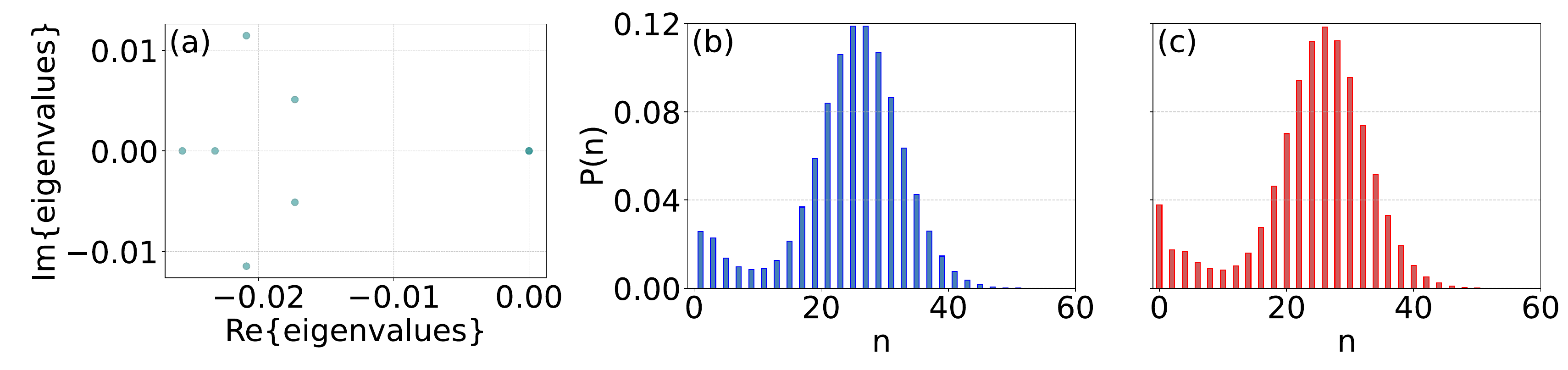}
        \caption{(a) The first 8 eigenvalues of the Liouvillian for $\omega_a$ = $\omega_c$ = 1, $N = 4$, $M = 50$, $\kappa_1$ = 0, $\kappa_2$ = 0.05 for $\lambda$ = 0.85. The Fock state population P(n) of the steady-state solution when the system was initialized at (b) $\ket{15}$ and (c) $\ket{10}$.}
        \label{fig: eigenvalue}
\end{figure}   

Figure~\ref{fig: eigenvalue}(a) shows the zoomed-in part of the Liouvillian spectrum close to the origin, for $N = 4$ and at $\lambda = 0.85$. As can be seen, there are two zero eigenvalues, i.e. a rank-2 kernel. Figure~\ref{fig: eigenvalue}(b) and (c) show the Fock state distribution of the NESS in two cases when the system is initialized in $\ket{15}$ and $\ket{10}$, respectively. 
\end{document}